\documentclass[letterpaper,twocolumn,showpacs,amsfonts,aps,%
                             superscriptaddress,prc,nofootinbib,floatfix]{revtex4}

\usepackage{fullpage}
\usepackage{amsmath}
\usepackage{bm}
\usepackage{graphicx}

\usepackage{color}
\usepackage[colorlinks=true, linkcolor=black, citecolor=blue, urlcolor=blue]{hyperref}

\newcommand{\beq}[1]{\begin{equation}\label{#1}}
\newcommand{\eeq}{\end{equation}}

\begin{document}

\title{Heavy-flavor dynamics in relativistic p-Pb collisions at $\sqrt{s_{NN}}=5.02$ TeV}

\author{Yingru Xu}
\email[Correspond to\ ]{yx59@phy.duke.edu}
\affiliation{Department of Physics, Duke University, Durham, NC 27708, USA}

\author{Shanshan Cao}
\affiliation{Nuclear Science Division, Laerence Berkeley National Laboratory, Berkeley, CA 94720, USA}

\author{Marlene Nahrgang}
\affiliation{Department of Physics, Duke University, Durham, NC 27708, USA}

\author{Weiyao Ke}
\affiliation{Department of Physics, Duke University, Durham, NC 27708, USA}

\author{Guang-You Qin}
\affiliation{Insititute of Particle Physics and Key Laboratory of Quark and Leption Physics (MOE), Central China Normal University, Wuhan, 430079, China}

\author{Jussi Auvinen}
\affiliation{Department of Physics, Duke University, Durham, NC 27708, USA}

\author{Steffen A. Bass}
\affiliation{Department of Physics, Duke University, Durham, NC 27708, USA}

\begin{abstract}
We investigate the heavy flavor dynamics in the quark-gluon plasma (QGP) medium created in p-Pb collisions at the CERN Large Hadron Collider (LHC). In the (3+1)-dimensional viscous hydrodynamics model describing QGP medium, the dynamics of heavy quarks are studied in an improved Langevin framework incorporating both collisional and radiative energy loss. The hadronization of the heavy quarks is given by a hybrid model of fragmentation and recombination. We find that the in-medium evolution of charm quarks raises the D-meson $R_{pPb}$ at low transverse momenta while it suppresses the D-meson $R_{pPb}$ at intermediate momenta. In addition, the elliptic flow of D-meson is calculated. For a diffusion coefficient which reproduces central $R_{AA}$ data at the LHC, we find a much smaller D-meson $v_2$ compared to the light hadrons. This observation indicates an incomplete coupling between heavy quarks with the medium, due to the reduced medium size compared to AA collisions.
\end{abstract}

\maketitle

\section{Introduction}
\label{S:1}

Heavy flavor, including charm and bottom quarks, serves as valuable hard probes for the quark-gluon plasma due to their predominant production in the early stage of the collision, their experience of the full evolution of the medium, and their incomplete thermalization during the propagation. The properties of the QGP medium can be probed by the study of the heavy meson nuclear modification factor $R_{AA}$ and their elliptic flow $v_2$. Those two observables are believed to be sensitive to the energy density and temperature of the medium, as well as its transport properties. At both LHC and RHIC, the nuclear modification factor $R_{AA}$ and the elliptic flow coefficients of D-mesons are observed to be comparable to light hadrons \cite{Abelev:2013lca}\cite{ALICE:2012ab}\cite{Adare:2010de}. This urges us to study the heavy flavor dynamics in more detail. 

In order to explore the properties of the QGP medium, measurements of heavy flavor observables in p-Pb collisions at LHC is believed to be a reference for interpreting the results of AA collisions. In addition, due to the reduced size of the nuclear medium as well as the short life time, the in-medium energy loss of heavy quarks in p-Pb collisions is expected to be less compared to AA collisions, which enables us to disentangle the cold nuclear matter (CNM) effects  from the hot nuclear matter (HNM) effects. Moreover, the observation of the ridge in pPb collisions and the comparable flow between high multiplicity p-Pb collisions and peripheral Pb-Pb collisions  \cite{CMS:2012qk}\cite{Chatrchyan:2013nka} indicate collectivity in p-Pb collisions, as well as applicability of hydrodynamical description of the medium created in p-Pb collisions \cite{Bozek:2014era}\cite{Bzdak:2013zma}\cite{Werner:2013tya}. 

\section{Dynamics of Heavy Flavor in Heavy-ion Collisions}
\label{S:2}

The full evolution in pA collisions can be separated into 3 stages: the initial state; the in-medium evolution; and the hadronization of heavy quarks into heavy mesons.

The initial state of both heavy quarks and the QGP medium is produced via a participant parton based entropy deposition model in position space, while a leading order pQCD is used to calculate the initial momentum distribution of heavy flavor. In the participant parton model, each parton is assigned a fluctuated thickness function:

\begin{equation}
T_{A,B} (x_{\perp})= \omega_{A,B} \int_{}^{} dz \rho_{A,B}(x_{\perp}, z)
\end{equation}

where $\omega_{A,B}$ are independent weight factors obtained from a gamma distribution. The rapidity dependence of the entropy density is inspired by a CGC formula in ref \cite{McLerran:2015lta}. In the pQCD calculation, the CTEQ and EPS09 parametrizations are adopted for the parton distribution function and the nuclear shadowing effect.

 The full space-time evolution of the QGP medium is described via the (3+1)-dimensional viscous hydrodynamics model vHLLE \cite{Karpenko:2013wva}. The ratio of shear viscosity over entropy density is chosen as $\eta/s$=0.08 and the initial state is tuned to produce the pseudorapidity distribution of charged hadrons comparable to the experimental results \cite{ALICE:2012xs}. An improved Langevin approach \cite{Cao:2015hia} is adopted to describe the in-medium propagation of heavy quarks. The Langevin approach incorporates both collisional energy loss and radiative energy loss:
\begin{equation}
\frac{d\vec{p}}{dt}=-\eta_D(p)\vec{p}+\vec{\xi}+\vec{f_g}
\end{equation}
The first two terms on the right hand side of the equation are the drag and thermal random forces which describe the quasi-elastic scattering between heavy quarks and light partons in the medium. The third term $\vec{f_g}$ is the recoil force induced due to the heavy quarks' emission of bremsstrahlung gluons. 

\begin{figure}[!tbp]
  \centering
  {\includegraphics[width=0.49\textwidth]{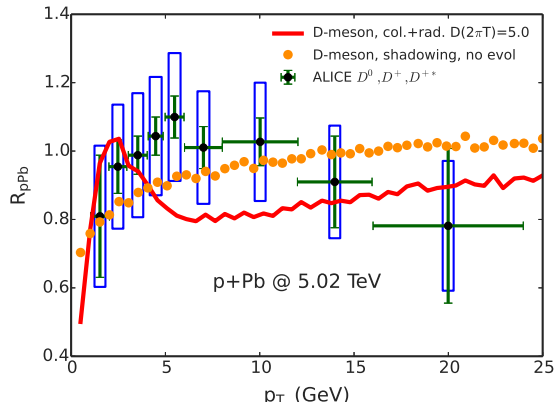}\label{f1a}}
  \hfill
  \caption{ Nuclear modification factor of D-mesons in p-Pb collisions. The $R_{pPb}$ is calculated under minimum bias situation with only nuclear shadowing effects (yellow dots), and shadowing effects as well as in-medium evolution (red line).}
\end{figure}

We utilize a higher-twist calculation \cite{Zhang:2003wk} for the radiated gluon distribution and relate the gluon transport coefficient $\hat{q}$ with the spatial diffusion coefficient D via $D=4 \frac{T^2}{\hat{q}}\frac{C_A}{C_F}$. A more detailed description can be found in \cite{Cao:2015hia}. Using this construction, the only free parameter in the improved Langevin approach is the spatial diffusion coefficient. It is usually determined by tuning D to the D-meson $R_{AA}$ in the most 0-7.5\% central Pb+Pb collisions at $\sqrt{s_{NN}}$ = 2.76 TeV \cite{Cao:2013ita}. In this study the value is taken as $D (2\pi T)$ = 5.0.

At hadronization, the particlization of the QGP matter is calculated according to the Cooper-Frye prescription \cite{Karpenko:2015xea}. For the heavy quarks, we utilize a hybrid model of fragmentation and recombination \cite{Fries:2003vb} to hadronize heavy quarks into heavy mesons. It has been shown \cite{Cao:2013ita} that while the fragmentation process is dominant for high momenta, the recombination of heavy quarks and light partons significantly enhances the production of heavy mesons at low and intermediate momenta.
\section{Nuclear Modification Factor and Flow}
\label{S:3}
\begin{figure}
\centering
  {\includegraphics[width=0.49\textwidth]{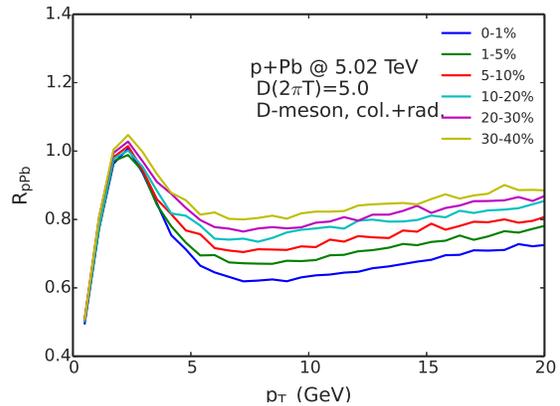}\label{f1b}
    \caption{ Nuclear modification factor of D-mesons for different centrality classes in p-Pb collisions. The $R_{pPb}$ is calculated with nuclear shadowing effects as well as in-medium evolution.}
     } 
\end{figure}

\begin{figure}[!tbp]
  \centering
  {\includegraphics[width=0.49\textwidth]{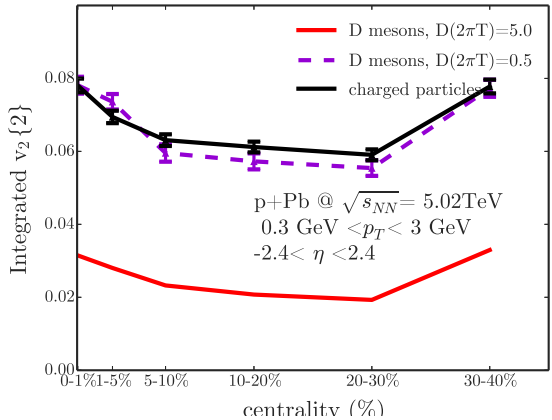}\label{f2a}}
  \hfill
  \caption{The $p_T$ integrated elliptic flow $v_2$\{2\} of charged particles (black line) and D-mesons (red line and purple dash line) as a function of centrality. The elliptic flow is calculated by Q cumulant method\cite{Bilandzic:2010jr}. The shear viscosity entropy ratio is chosen as $\eta/s$= 0.08, the diffusion coefficients is chosen as $D(2 \pi T)$ = 0.5 (purple dashed line) and $D(2 \pi T)$ = 5.0 (red line). The $p_T$ and $\eta$ cut for both charged particles and D-mesons are 0.3 GeV $< p_T <$ 3.0 GeV,  $\left| \eta \right| <$ 2.4.}
\end{figure}
Our results for the nuclear modification factor are shown in Fig.1 and Fig.2. Fig.1 compares the nuclear modification factor of D-meson $R_{pPb}$ including only the initial nuclear shadowing to that obtained from the full evolution. Within the error bars the experimental data \cite{Abelev:2014hha} is consistent with unity with some indications for an enhancement at around $p_T=$ 5 GeV and a suppression towards higher $p_T$. Including only the initial nuclear shadowing but no in-medium evolution for the heavy quark, the $R_{pPb}$ is depleted at small $p_T$ and reach unity at higher $p_T$. If we take the in-medium evolution into account as well, we observe that the in-medium evolution raises the $R_{pPb}$ at low momenta and further suppresses the production of D-meson at intermediate momenta. While Fig.1 shows results from minimum bias, Fig.2 presents the $R_{pPb}$ for different centrality classes. The centrality classes are sliced according to the total entropy density of each event (among 1 million random events) generated by the initial condition. As expected, the suppression at intermediate and high $p_T$ increases from peripheral to most central collisions. The centrality dependence of $R_{pPb}$ of D-meson can therefore serve as a useful observable to distinguish between CNM and HNM effects.

\begin{figure}
\centering
  {\includegraphics[width=0.49\textwidth]{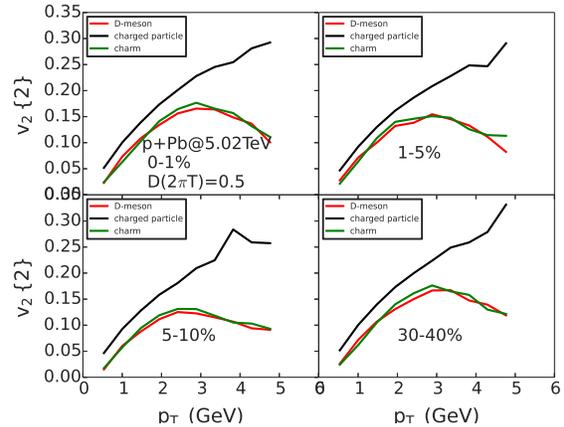}\label{f2b}
  \caption{The $p_T$ differential elliptic flow $v_2$\{2\} of charged particles (black line), charm quarks (green line) and D-mesons (red line) in different centrality classes. The diffusion coefficient is $D(2 \pi T)$ = 0.5, as it gives the best description of $v_2$ in Pb-Pb collisions at 30-50\% centrality. }
  }
\end{figure}
The results for the elliptic flow are presented in Fig.3 and Fig.4. In Fig.3 we show the results for two diffusion coefficients: $D=5.0/2\pi T$ and $D=0.5/2\pi T$. The former one gives the best description of D-meson $R_{AA}$ in Pb-Pb collisions in 0-7.5\% centrality while the later one gives the best description of D-meson $v_2$ in Pb-Pb collisions in 30-50\% centrality. This discrepancy shows that our Langevin approach to the in-medium dynamics cannot reproduce both the $R_{AA}$ and the $v_2$ consistently. This is currently under investigation. For $D=5.0/2\pi T$, a significantly suppressed elliptic flow of D-mesons indicates an incomplete coupling between the charm quarks and the medium in the p-Pb collisions. This is due to the short life time and small size of QGP medium. On the other hand, for $D=0.5/2\pi T$ case, the D-meson integrated $v_2$ is similar to that of the charged particles in both shape and magnitude. It should be mentioned, however, that such a small diffusion coefficient is strongly disfavored by lattice QCD \cite{Banerjee:2011ra}\cite{Ding:2012sp}\cite{Kaczmarek:2014jga} and model calculations. Fig.4 shows the $p_T$-differential flow for charged particles, charm quarks and D-mesons for four different centralities. The flows of D-mesons and charm quarks are comparable with each other, indicating that hadronization via recombination only has a small contribution to the final D-meson $v_2$.

\section{Conclusion}
\label{S:4}
In this work we have investigated the medium modification of heavy quarks in the hot medium created in p-Pb collisions at $\sqrt{s_{NN}}=5.02$ TeV at LHC. The aim of this study was to consistently describe the light and soft bulk evolution and the dynamics of heavy flavor.  We presented the results of nuclear modification factor of D-mesons in the p-Pb collisions in the minimum bias criterion and for different centrality classes. Although he accuracy of current data does not allow us to distinguish unambiguously between scenarios of only CNM or CNM + HNM effects if the pA results are viewes independently of the AA results, the centrality dependence of the D-meson $R_{pPb}$ can help to disentangle the contributions from CNM and HNM effects. In addition, we presented our calculation of elliptic flow of D-mesons in p-Pb collisions that are indicative of an incomplete coupling between the charm quarks with the QGP medium. 

\section{Acknowledgements}
We thank I. Karpenko for providing the (3+1) dimensional relativistic viscous hydrodynamics code vHLLE. Computational resources were provided by the Open Science Grid (OSG). This work has been supported by the U.S Department of Energy under grant DE-FG02-05ER41367 and DE-AC02- 05CH11231. M.N. acknowledges support from a fellowship within the Postdoc-Program of the German Academic Exchange Service (DAAD). G.-Y.Q. acknowledges the support from the Natural Science Foundation of China (NSFC) under Grant No. 11375072.  \\
\\


\begin{thebibliography}{99}

\bibitem{Abelev:2013lca} 
  B.~Abelev {\it et al.} [ALICE Collaboration],
  Phys.\ Rev.\ Lett.\  {\bf 111}, 102301 (2013)
  [arXiv:1305.2707 [nucl-ex]].
  
\bibitem{ALICE:2012ab} 
  B.~Abelev {\it et al.} [ALICE Collaboration],
  JHEP {\bf 1209}, 112 (2012)
  [arXiv:1203.2160 [nucl-ex]].

\bibitem{Adare:2010de} 
  A.~Adare {\it et al.} [PHENIX Collaboration],
  Phys.\ Rev.\ C {\bf 84}, 044905 (2011)
  [arXiv:1005.1627 [nucl-ex]].

\bibitem{CMS:2012qk} 
  S.~Chatrchyan {\it et al.} [CMS Collaboration],
  Phys.\ Lett.\ B {\bf 718}, 795 (2013)
  [arXiv:1210.5482 [nucl-ex]].
 
 \bibitem{Chatrchyan:2013nka} 
   S.~Chatrchyan {\it et al.} [CMS Collaboration],
   Phys.\ Lett.\ B {\bf 724}, 213 (2013)
   [arXiv:1305.0609 [nucl-ex]].
   
\bibitem{Bozek:2014era} 
  P.~Bozek and W.~Broniowski,
  Nucl.\ Phys.\ A {\bf 926}, 16 (2014)
  [arXiv:1401.2367 [nucl-th]].
  
\bibitem{Bzdak:2013zma} 
  A.~Bzdak, B.~Schenke, P.~Tribedy and R.~Venugopalan,
  Phys.\ Rev.\ C {\bf 87}, no. 6, 064906 (2013)
  [arXiv:1304.3403 [nucl-th]].

\bibitem{Werner:2013tya} 
  K.~Werner, B.~Guiot, I.~Karpenko and T.~Pierog,
  Phys.\ Rev.\ C {\bf 89}, no. 6, 064903 (2014)
  [arXiv:1312.1233 [nucl-th]].
    
\bibitem{McLerran:2015lta} 
   L.~McLerran and M.~Praszalowicz,
   arXiv:1507.05976 [hep-ph].

\bibitem{Karpenko:2013wva} 
  I.~Karpenko, P.~Huovinen and M.~Bleicher,
   Comput.\ Phys.\ Commun.\  {\bf 185}, 3016 (2014)
  [arXiv:1312.4160 [nucl-th]].
  
\bibitem{ALICE:2012xs} 
  B.~Abelev {\it et al.} [ALICE Collaboration],
  Phys.\ Rev.\ Lett.\  {\bf 110}, no. 3, 032301 (2013)
  [arXiv:1210.3615 [nucl-ex]].
  
\bibitem{Cao:2015hia} 
  S.~Cao, G.~Y.~Qin and S.~A.~Bass,
  Phys.\ Rev.\ C {\bf 92}, no. 2, 024907 (2015)
  [arXiv:1505.01413 [nucl-th]].

 \bibitem{Zhang:2003wk} 
   B.~W.~Zhang, E.~Wang and X.~N.~Wang,
   Phys.\ Rev.\ Lett.\  {\bf 93}, 072301 (2004)
   [nucl-th/0309040].



\bibitem{Cao:2013ita} 
  S.~Cao, G.~Y.~Qin and S.~A.~Bass,
  Phys.\ Rev.\ C {\bf 88}, no. 4, 044907 (2013)
  [arXiv:1308.0617 [nucl-th]].
  
\bibitem{Karpenko:2015xea} 
  I.~A.~Karpenko, P.~Huovinen, H.~Petersen and M.~Bleicher,
  Phys.\ Rev.\ C {\bf 91}, no. 6, 064901 (2015)
  [arXiv:1502.01978 [nucl-th]].
  
\bibitem{Fries:2003vb}
   R. J. Fries, B. Muller, C. Nonaka and S.~A.~Bass,
   Phys.\ Rev.\ Lett.\  {\bf 90} (2003) 202303
   [nucl-th/0301087].
     

 
\bibitem{Qin:2013bha} 
  G.~Y.~Qin and B.~Müller,
  Phys.\ Rev.\ C {\bf 89}, no. 4, 044902 (2014)
  [arXiv:1306.3439 [nucl-th]].

 
 
\bibitem{Abelev:2014hha} 
  B.~B.~Abelev {\it et al.} [ALICE Collaboration],
  Phys.\ Rev.\ Lett.\  {\bf 113}, no. 23, 232301 (2014)
  [arXiv:1405.3452 [nucl-ex]].
  
\bibitem{Cao:2014dja} 
  S.~Cao, G.~Y.~Qin and S.~A.~Bass,
  Nucl.\ Phys.\ A {\bf 931}, 569 (2014)
  [arXiv:1408.0503 [nucl-th]].

\bibitem{Bilandzic:2010jr} 
  A.~Bilandzic, R.~Snellings and S.~Voloshin,
  Phys.\ Rev.\ C {\bf 83}, 044913 (2011)
  [arXiv:1010.0233 [nucl-ex]].
   
\bibitem{Banerjee:2011ra} 
  D.~Banerjee, S.~Datta, R.~Gavai and P.~Majumdar,
  Phys.\ Rev.\ D {\bf 85}, 014510 (2012)
  [arXiv:1109.5738 [hep-lat]].
 
 \bibitem{Ding:2012sp} 
   H.~T.~Ding, A.~Francis, O.~Kaczmarek, F.~Karsch, H.~Satz and W.~Soeldner,
   Phys.\ Rev.\ D {\bf 86}, 014509 (2012)
   [arXiv:1204.4945 [hep-lat]].
  
\bibitem{Kaczmarek:2014jga} 
  O.~Kaczmarek,
  Nucl.\ Phys.\ A {\bf 931}, 633 (2014)
  [arXiv:1409.3724 [hep-lat]].

\end{thebibliography}
\end{document}